# Precipitable water vapour measurement using GNSS data in the Atacama Desert for millimetre and submillimetre astronomical observations

Junna Sugiyama[1]★, Haruki Nishino[2]★† and Akito Kusaka[1,2,3,4]

[1]*Department of Physics, The University of Tokyo, Tokyo 113-0033, Japan*
[2]*Research Center for the Early Universe (RESCEU), Graduate School of Science, The University of Tokyo, Tokyo 113-0033, Japan*
[3]*Physics Division, Lawrence Berkeley National Laboratory, Berkeley, CA 94720, USA*
[4]*Kavli Institute for the Physics and Mathematics of the Universe (WPI), The University of Tokyo, Chiba 277-8583, Japan*

**ABSTRACT**
Precipitable water vapour (PWV) strongly affects the quality of data obtained from millimetre- and submillimetre-wave astronomical observations, such as those for cosmic microwave background measurements. Some of these observatories have used radiometers to monitor PWV. In this study, PWV was measured from 2021 April to 2022 April using Global Navigation Satellite System (GNSS) instruments in the Atacama Desert, Chile, where several millimetre- and submillimetre-wave telescopes are located. We evaluated the accuracy of these measurements by comparing them to radiometer measurements. We calculated the PWV from GNSS data using CSRS-PPP (Canadian Spatial Reference System Precise Point Positioning), an online software package. When using GNSS data alone, the estimated PWV showed a systematic offset of $+1.08$ mm. When combining GNSS data with data from a barometer, which was co-located with the GNSS receiver, the estimated PWV showed a lower systematic offset of $-0.05$ mm. The GNSS PWV showed a statistical uncertainty of 0.52 mm with an averaging time of an hour. Compared to other PWV measurement methods, GNSS instruments are robust in bad weather conditions, have sufficient time resolution, and are less expensive. By demonstrating good accuracy and precision in low-PWV conditions, this paper shows that GNSS instruments are valuable tools for PWV measurements for observing site evaluation and data analysis for ground-based telescopes.

## 1 INTRODUCTION

Water vapour in the atmosphere strongly absorbs electromagnetic waves in millimetre and submillimetre wavelengths, and its radiation affects astronomical observations. Precipitable water vapour (PWV) is a key parameter in evaluating the quality of the observational data. Ground-based astronomical observations, such as those for the cosmic microwave background (CMB), require excellent atmospheric conditions and accurate monitoring of PWV.

The Atacama Desert in Chile is known for its arid climate with low PWV (Radford & Holdaway 1998; Lay & Halverson 2000; Giovanelli et al. 2001; Bustos et al. 2014; Cortés et al. 2020). This makes it an ideal location for ground-based millimetre/submillimetre observatories, such as the Atacama Submillimeter Telescope Experiment (Ezawa et al. 2004), NANTEN2 (Onishi et al. 2008), Atacama Pathfinder Experiment (APEX; Güsten et al. 2006), and Atacama Large Millimeter/Submillimeter Array (ALMA; Wootten & Thompson 2009). In particular, Cerro Toco hosts several CMB experiments, including the Atacama Cosmology Telescope (Swetz et al. 2011), the Atacama B-mode Search (ABS; Kusaka et al. 2018), the POLARBEAR experiment (Kermish et al. 2012), the Simons Array (SA; Suzuki et al. 2016), and the Cosmology Large Angular Scale Surveyor (Essinger-Hileman et al. 2014). Planned experiments for this location include the Simons Observatory (Ade et al. 2019) and CMB-S4 (Abazajian et al. 2019). CMB experiments use PWV measurements for the selection of observation sites and data analysis. Since these experiments have observation time-scales of several years, the PWV estimation method must be robust and continuous over years.

The Global Navigation Satellite System (GNSS) is one of the tools that are used to measure PWV. GNSS is a general term used to describe satellite-based positioning systems such as Global Positioning System (GPS), Global'naya Navigatsionnaya Sputnikovaya Sistema (GLONASS), Galileo, BeiDou, Navigation with Indian Constellation (NavIC), and Quasi-Zenith Satellite System (QZSS). Estimation of PWV using GPS was first reported in 1992 (Bevis et al. 1992), and a number of studies have shown that GNSS data can be used to estimate PWV as accurately as other methods (e.g. Schneider et al. 2010). Estimating PWV in dry climates using GNSS data is known to be more challenging than in humid climates, but papers have reported successful estimations in dry regions such as Antarctica (Suparta et al. 2007, 2009; Vázquez B & Grejner-Brzezinska 2013; Negusini et al. 2016, 2021; Ding, Chen & Tang 2022), Northern Sweden (Buehler et al. 2012), the Himalayas (Jade et al. 2004; Jade & Vijayan 2008; Joshi et al. 2013; Ningombam et al. 2016; Ningombam, Jade & Shrungeshwara 2018), and Canary Islands (García-Lorenzo et al. 2010; Schneider et al. 2010).

★ E-mail: junna.sugiyama@phys.s.u-tokyo.ac.jp (JS);
haruki.nishino@cmb.phys.s.u-tokyo.ac.jp (HN)
† Present address: Japan Synchrotron Radiation Research Institute (JASRI), 1-1-1, Kouto, Sayo-cho, Sayo-gun, Hyogo 679-5198, Japan.

There are several other instruments that can measure PWV, such as radiometers, radiosondes, and the Moderate Resolution Imaging Spectroradiometer (MODIS) aboard satellites. However, instruments that can continuously and accurately observe PWV are limited. Radiometer measurements are easily distorted by the presence of liquid water and may significantly overestimate PWV (Shangguan et al. 2015). Observations should be excluded around precipitation events to avoid this error, but this results in periods of downtime. Radiosondes are expensive to operate for long periods of time. MODIS can only measure PWV twice per site per day and is known to overestimate PWV in dry weather (Li, Muller & Cross 2003; Vaquero-Martínez et al. 2017; He & Liu 2019; Sam Khaniani, Nikraftar & Zakeri 2020).

In contrast, GNSS instruments are inexpensive these days and stable in bad weather conditions, making them suitable for supporting observatories. Wood-Vasey, Perrefort & Baker (2022) used GPS measurements of PWV to support photometric calibrations in the red optical and near-infrared wavelengths in their optical observations. In this study, GNSS measurements of PWV were evaluated to support millimetre/submillimetre astronomical observatories, especially CMB experiments. Using GNSS data, continuous estimates of PWV at Cerro Toco over a year with sufficient accuracy and sampling rate are presented.

This paper is organized as follows. Section 2 describes the calculation of PWV from GNSS data. Section 3 describes the instruments, location, and systems used to collect data. Section 4 describes the software and data processing flow for the PWV estimation. Section 5 discusses the performance of the GNSS-derived PWV compared to the radiometer. Section 6 discusses the uncertainties of the GNSS-derived PWV. Section 7 discusses the quality of GNSS-derived PWV estimates in the context of data analysis and site evaluation for CMB observatory. Section 8 summarizes the results and conclusions.

## 2 METHODOLOGY

The method of deriving PWV information from GPS data was established by Bevis et al. (1992). As electromagnetic signals from satellites propagate through the troposphere, they are delayed, which depends on the index of refraction along the propagation path. The tropospheric delay in the zenith direction is called the zenith total delay (ZTD) and is expressed in metres. ZTD is estimated by observing dual frequency signals from multiple GNSS satellites. ZTD can be written as the sum of the zenith hydrostatic delay (ZHD) and the zenith wet delay (ZWD; Davis et al. 1985):

$$\mathrm{ZTD} = \mathrm{ZHD} + \mathrm{ZWD}. \quad (1)$$

ZHD is the dry component of the delay, while ZWD is a wet component and is proportional to the PWV. ZHD is typically of the order of 1.25 m on the Atacama Desert and accounts for almost all of ZTD. ZHD is accurately modelled using meteorological parameters. ZWD is estimated through two steps: ZTD measurement from GNSS data and ZHD estimation from meteorological parameters. Since ZWD is of the order of millimetres and varies rapidly, it has not been modelled as well as ZHD.

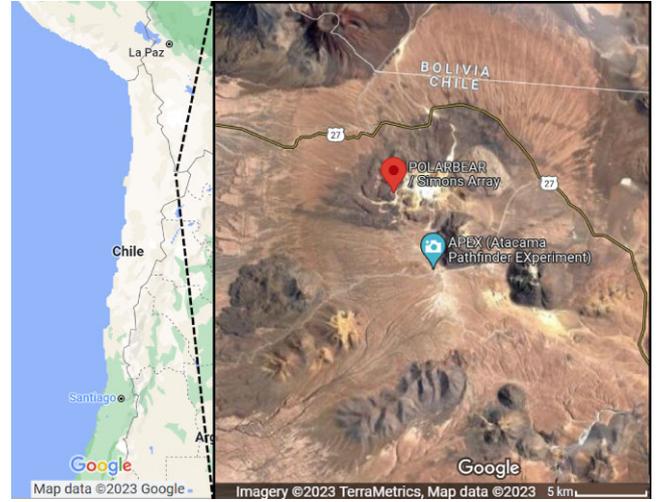

**Figure 1.** A map of the area around the SA experiment (Copyright 2023 Google; TerraMetrics, LLC – www.terrametrics.com). The radiometer data at the APEX site were used for the reference values of PWV.

ZHD is modelled by Davis et al. (1985) as a function of the surface pressure $P$:

$$\mathrm{ZHD} = 10^{-6} \frac{k_1 R_\mathrm{d}}{g_m} P \ (m), \quad (2)$$

where $k_1 = 77.604$ K hPa$^{-1}$ is the refractivity constant for dry air, $R_\mathrm{d} = 287.04$ J K$^{-1}$ kg$^{-1}$ is the specific gas constant for dry air, and $g_m$ is the mean gravity:

$$g_m = \frac{\int \rho(z) g(z) \mathrm{d}z}{\int \rho(z) \mathrm{d}z}, \quad (3)$$

where $\rho(z)$ and $g(z)$ are air density and gravity acceleration, respectively, as a function of zenith height $z$. We estimate $g_m$ with the Bosser et al. (2007) model parametrized by latitude, surface elevation, and time.

ZWD is converted to PWV as follows:

$$\mathrm{PWV} = \Pi \times \mathrm{ZWD}. \quad (4)$$

The conversion factor $\Pi$ is known to have an approximate value of $1/6.5 \sim 0.15$ (Bevis et al. 1992; Rocken et al. 1993). More accurately, $\Pi$ is estimated as a function of the weighted mean temperature $T_m$:

$$\Pi(T_m) = \frac{10^6}{\rho R_v (k_3/T_m + k_2')}, \quad (5)$$

where $\rho \sim 1000$ (kg m$^{-3}$) is the density of liquid water, $R_v \sim 461$ (J kg$^{-1}$ K$^{-1}$) is the specific gas constant for water vapour, and $k_3 \sim 3.7 \times 10^5$ (K$^2$ hPa$^{-1}$) and $k_2' \sim 22$ (K hPa$^{-1}$) are physical constants related to atmospheric refractivity (Askne & Nordius 1987; Bevis et al. 1994). $\Pi$ is expected to vary only by 4 per cent in the Atacama Desert. Thus, for simplicity, $\Pi$ is considered constant at its typical value of $\Pi = 0.151$ in this research. Alternatively, $T_m$ at each time can be estimated from our data set, as discussed in Appendix A.

## 3 MEASUREMENT SYSTEM

### 3.1 GNSS antenna and weather station

GNSS data were acquired at the SA experiment site at Cerro Toco (22°57′S, 67°47′W, 5200 m a.s.l.). Fig. 1 shows a map of the SA

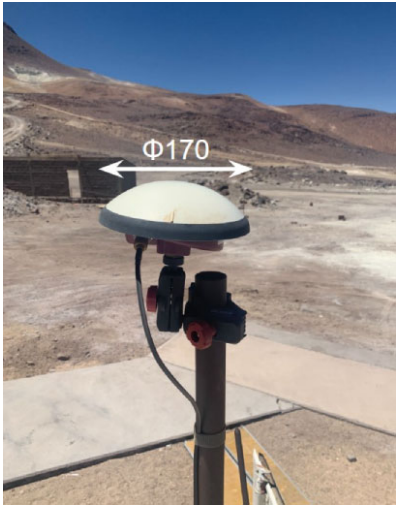

**Figure 2.** GNSS antenna (Tallysman VSP6037L) installed at the SA site.

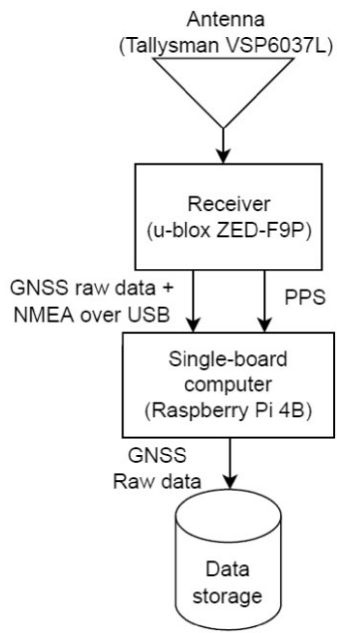

**Figure 3.** Block diagram of the GNSS antenna and receiver system.

experiment site. A GNSS antenna (Tallysman VSP6037L[1]) was installed on an intermodal container in 2021 April, as shown in Fig. 2. The surface pressure was measured by a barometer on a weather station (Davis Vantage Pro 2[2]) co-located with the GNSS antenna. It outputs the surface pressure every minute with a resolution of 0.1 hPa and accuracy of 1.0 hPa, and its range is 540–1100 hPa. A small percentage of the data fell below this range.

### 3.2 GNSS receiver system

Fig. 3 shows a block diagram of the GNSS receiver system. The raw GNSS signal from the antenna is transmitted through a coaxial cable to a GNSS receiver inside the equipment container. A multiband GNSS receiver (u-blox ZED-F9P[3]) was employed, which supports the major GNSS systems, including GPS, GLONASS, Galileo, and BeiDou. In this paper, we only used GPS and GLONASS observation data for the analysis.

The receiver was configured to output raw GNSS data for later post-processing with software. A single-board computer (Raspberry Pi 4B[4]) receives raw GNSS data from the receiver module and streams them to a data storage computer using the Transmission Control Protocol (TCP) server functionality of rtkrcv in the RTKLIB[5] software package. Raw GNSS data are converted to the Receiver Independent Exchange (RINEX) format using convbin in the RTKLIB package. The receiver was also configured to output National Marine Electronics Association (NMEA) messages and a pulse-per-second signal so that the single-board computer can synchronize its system clock to GNSS-referenced time.

## 4 ANALYSIS

### 4.1 Precise point positioning

Precise point positioning (PPP) is one of the GNSS analysis techniques for achieving fine-tuned positioning in the centimetre or millimetre range. PPP does not require nearby reference stations, unlike other GNSS positioning techniques such as real-time kinematic or differential GNSS. Instead, the precise satellite products, which correct the satellite orbit and clock errors, are used to assist in the positioning. Our ZTD estimation was performed using CSRS-PPP (Canadian Spatial Reference System PPP), a free online PPP service, since it has high accuracy compared to other software (Guo 2005; Mendez Astudillo et al. 2018; Vázquez-Ontiveros et al. 2023).

The CSRS-PPP overview is described in Tétreault et al. (2005), and recent updates are provided on its website.[6] CSRS-PPP computes ZTD from uploaded GNSS data in RINEX format along with the precise satellite products from the International GNSS Service and Natural Resources Canada (NRCan). Among their 'ultra-rapid', 'rapid', and 'final' data product options, 'final' products are used in the calculations. Only GPS and GLONASS observations are processed, as other satellites are not yet supported by this software. Satellites with elevations less than $7.5°$ are also excluded from the analysis to reduce multipath errors.

CSRS-PPP also calculates ZHD using gridded Vienna Mapping Function 1 (VMF1). Therefore, it is capable of estimating PWV from GNSS data alone. However, as discussed in Section 5, the accuracy of the PWV estimation can be improved with data from a barometer co-located with the GNSS instrument.

### 4.2 PWV estimations at the SA site

Fig. 4 illustrates how ZWD was derived. The raw GNSS data were converted to RINEX observation files and uploaded to the CSRS-PPP website. CSRS-PPP outputs both ZTD timestreams from the GNSS data and ZHD timestreams from the gridded VMF1. ZHD was also calculated using barometer data at the SA site and equation (2).

---

[1] https://www.tallysman.com/product/vsp6037l-verostar-full-gnss-antenna-l-band
[2] https://www.davisinstruments.com/pages/vantage-pro2
[3] https://www.u-blox.com/en/product/zed-f9p-module
[4] https://www.raspberrypi.com/products/
[5] https://www.rtklib.com/
[6] https://webapp.csrs-scrs.nrcan-rncan.gc.ca/geod/tools-outils/ppp-update.php

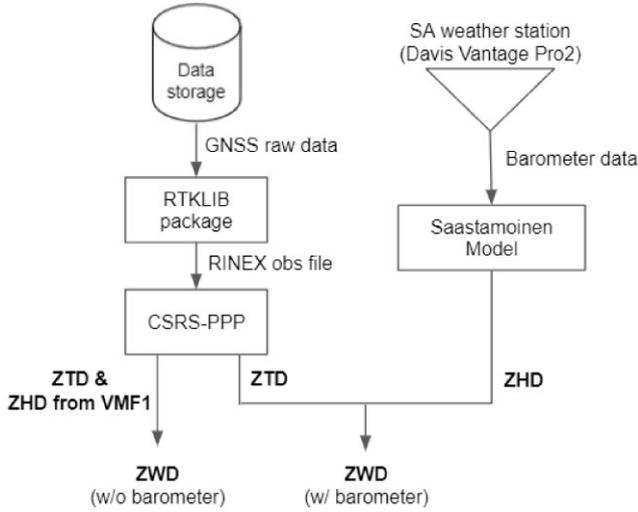

**Figure 4.** Block diagram of the ZWD estimation using GNSS with and without barometer data.

Thus, ZWD was calculated by two methods: one using the barometer-derived ZHD (denoted as '*w/ barometer*') and the other using VMF1-derived ZHD (denoted as '*w/o barometer*'). ZWD was converted to PWV using equation (4).

### 4.3 Radiometer data for comparison

To evaluate the accuracy of the GNSS-derived PWV measurements, they were compared with PWV measurements obtained by a ground-based microwave radiometer that is used in APEX.[7] Their estimated systematic uncertainty is less than 3 per cent (Cortés et al. 2020).

The distances between APEX and SA are 6 km horizontally and 150 m vertically, which lead to statistical and systematic differences in PWV, respectively. In the Atacama Desert, the spatial inhomogeneity of PWV causes a path-length variation of the order of 200–300 μm between points that are 6 km apart, with an average time of 10 s (ALMA Partnership 2015; Matsushita et al. 2017). This corresponds to PWV variation of the order of 0.05 mm, assuming $\Pi = 0.15$. This is a small enough difference that can be ignored in this study, as discussed in Section 5. APEX is situated at an altitude of 5050 m, while the SA experiment is at 5200 m. We correct for this vertical difference using the relationship given by Otárola et al. (2010):

$$\mathrm{PWV_{RM}} = \exp\left(-\frac{5200 - 5050}{2300}\right) \mathrm{PWW_{APEX}}, \quad (6)$$

where $\mathrm{PWV_{APEX}}$ is the PWV measured by the APEX radiometer, and $\mathrm{PWV_{RM}}$ is the estimated radiometer PWV value for the SA site.

## 5 RESULTS

The accuracy of our GNSS-derived PWV ($\mathrm{PWV_{GNSS}}$) was evaluated through comparison to the radiometer-derived PWV ($\mathrm{PWV_{RM}}$). The analysis was done using 1 yr of GNSS data at the SA experimental site from 2021 April 18 to 2022 April 30.

Fig. 5 shows the time series of the *w/ barometer* $\mathrm{PWV_{GNSS}}$ and $\mathrm{PWV_{RM}}$. The 34 per cent of APEX radiometer data are missing because the radiometer shutter closes when bad weather is predicted, as well as during austral summer from January to March every year. The GNSS PWV data cover almost a full year aside from where the pressure data are missing or reached the lower limit of the barometer range (3 per cent in total). This downtime can be further reduced by adopting a barometer with a wider range. The $\mathrm{PWV_{GNSS}}$ follows the trend of $\mathrm{PWV_{RM}}$ while having nearly zero downtime.

---

[7]This publication is based on data acquired with the APEX. APEX is a collaboration between the Max-Planck-Institut fur Radioastronomie, the European Southern Observatory, and the Onsala Space Observatory.

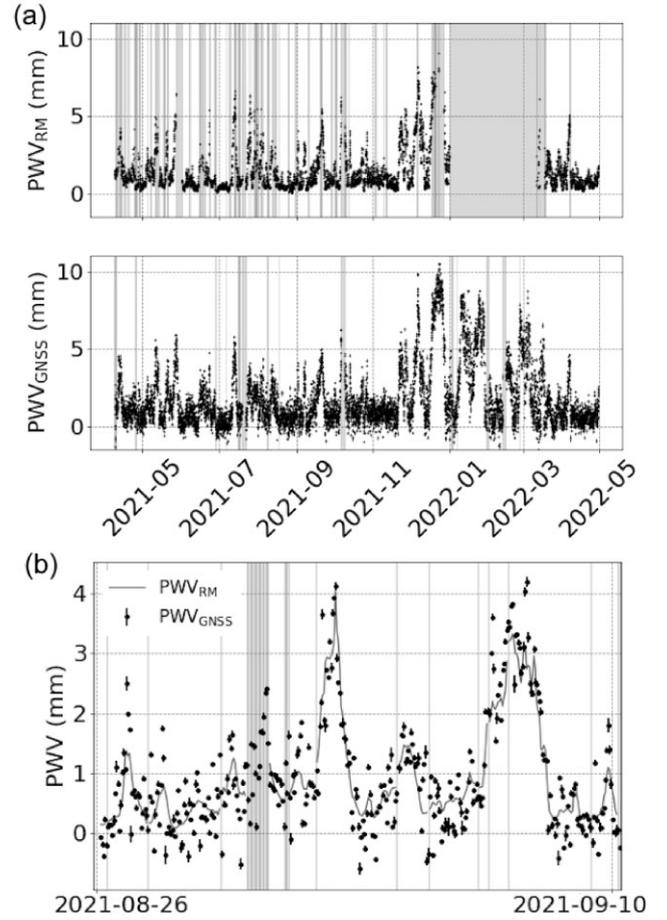

**Figure 5.** Hourly PWV time series. (a) PWV from radiometer estimate (top) and GNSS estimate (bottom). Time periods' missing data for more than 5 h are shown in bands. The radiometer data are missing when bad weather is predicted, as well as the shutdown period from January to March. The missing data from GNSS PWV occur where the barometer data are lost. (b) Time series of radiometer PWV and GNSS PWV over 15 d. Missing radiometer data are shown in grey bands. No GNSS PWV data are lost in this time range.

Fig. 6 shows a comparison of $\mathrm{PWV_{GNSS}}$ and $\mathrm{PWV_{RM}}$. As described in Section 4.2, two ZHD estimation methods were used to derive GNSS PWV. The left panel shows the *w/ barometer* $\mathrm{PWV_{GNSS}}$, while the right panel shows the *w/o barometer* $\mathrm{PWV_{GNSS}}$. We fit the linear relation between $\mathrm{PWV_{RM}}$ and $\mathrm{PWV_{GNSS}}$ by the following equation:

$$\mathrm{PWV_{GNSS}} = a\,\mathrm{PWV_{RM}} + b. \quad (7)$$

The fit parameters are $a = 0.996 \pm 0.006$ and $b = -0.05 \pm 0.01$ for *w/ barometer*, and $a = 1.046 \pm 0.006$ and $b = 1.08 \pm 0.01$ for *w/o barometer*.

The coefficient $a$ is close to unity in both cases, which justifies the use of $\Pi = 0.151$. The difference from unity is well within the uncertainties in $\mathrm{PWV_{RM}}$ (Section 4.3) and $\Pi$ (Appendix A). Using the barometer data reduces the offset $b$ from $+1.08$ to $-0.05$ mm. We consider that the on-site pressure data resulted in the accurate ZHD estimate, hence the small offset. Hereafter, *w/ barometer* PWV$_{GNSS}$ is just denoted as PWV$_{GNSS}$.

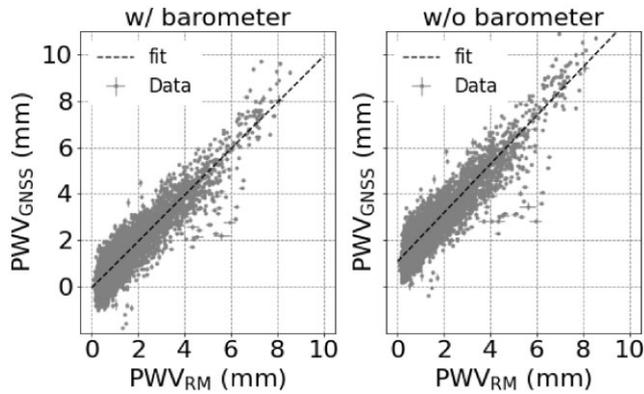

**Figure 6.** Scatter plot of hourly PWV measured by GNSS versus PWV measured by the radiometer. Left: *w/ barometer* PWV$_{GNSS}$ versus PWV$_{RM}$. Right: *w/o barometer* PWV$_{GNSS}$ versus PWV$_{RM}$. Using the co-located barometer makes the offset small. The time range where the data are lost, shown in Fig. 5, is not included in this plot.

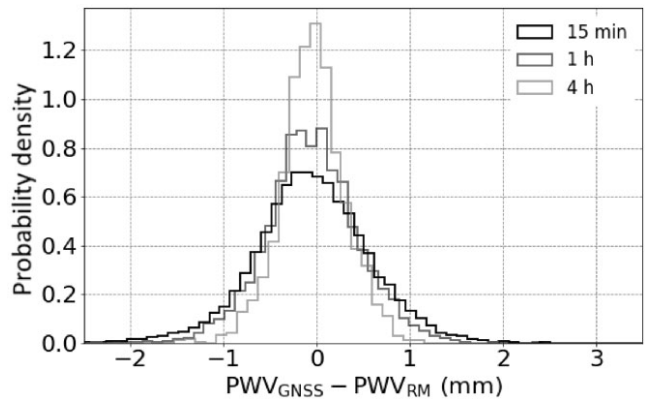

**Figure 7.** Histogram of PWV$_{GNSS}$ − PWV$_{RM}$ with average time of 15 min, 1 h, and 4 h. The time range where the data are lost, shown in Fig. 5, is not included in this plot.

Fig. 7 shows the distribution of PWV$_{GNSS}$ − PWV$_{RM}$ for different time averages of the PWV timestreams (15 min, 1 h, and 4 h). The statistical fluctuation is almost independent of the PWV values. The standard deviation (STD) of PWV$_{GNSS}$ − PWV$_{RM}$ is 0.64, 0.52, and 0.37 mm for the average time of 15 min, 1 h, and 4 h, respectively. These are dominated by the statistical uncertainty of PWV$_{GNSS}$, since the statistical uncertainty of PWV$_{RM}$ is less than 0.05 mm (see Section 4.3).

## 6 DISCUSSIONS FOR THE GNSS PWV UNCERTAINTY

We evaluated the GNSS PWV uncertainty in Section 5. Here, we discuss the source of uncertainties of our GNSS PWV, and the accuracy of our GNSS PWV in comparison to the literature.

The systematic uncertainty of GNSS PWV is represented by the deviation of *a* from 1 and the deviation of *b* from 0. It is expected to result from instrument miscalibration and modelling inaccuracies. The barometer miscalibration of 1 hPa affects *b* by 0.3 mm. We also found that the choice of the ZHD estimation model described in Section 2 affects *b* by about 0.1 mm. We used the $g_m$ estimation model of Bosser et al. (2007) and obtained $b = −0.05$ mm as shown in Section 5. For example, using the Saastamoinen (1973) model instead changed *b* to −0.14 mm. The altitude correction of the radiometer PWV with equation (6) changed *a* from 0.945 to 0.996.

The statistical uncertainty of GNSS PWV is represented by the STD of PWV$_{GNSS}$ − PWV$_{RM}$. It is expected to arise from the uncertainties in the pressure and ZTD measurements. The statistical uncertainty in the ZTD analysis is caused by the multipath error, the imperfection of the GNSS clock and orbit corrections, the ionosphere correction error, and the hardware noise. A pressure measurement uncertainty of 1 hPa would map to a PWV uncertainty of 0.3 mm, and a ZTD measurement uncertainty of 1 mm would map to a PWV uncertainty of 0.2 mm.

The uncertainties of our GNSS PWV are compared to other studies in Table 1. In the table, STD represents the standard deviation of the difference between the GNSS PWV and the PWV estimated by the method being compared. Although rigorous comparison is difficult due to differences in PWV ranges between studies, this study has the least uncertainty by comparing the raw values of STD.

## 7 APPLICATIONS TO CMB EXPERIMENTS

PWV is used to assess the data quality in the CMB observations. Many of modern ground-based CMB experiments use transition edge sensor (TES) bolometer arrays for their detectors. High optical loading from high PWV may make a TES inoperable, or may cause significant non-linearity of TES, which leads to systematic uncertainties (see e.g. Takakura et al. 2017; Appel et al. 2022). In order to improve data quality, some CMB experiments remove data where PWV is higher than a threshold value, typically around 3 mm for experiments optimized for the Atacama Desert. For example, POLARBEAR experiment discards its observations if the PWV exceeds a threshold of 4 mm (The POLARBEAR Collaboration 2017), and ABS experiment used a threshold of 2.5 mm (Kusaka et al. 2018). PWV measurements are also used in the instrument characterization, calibration, and null testing for data validation (Essinger-Hileman et al. 2016; Simon et al. 2016; Qu et al. 2023). In the null testing, the CMB observation data are split into two subsets based on an environmental or instrumental variable that may influence systematics, and consistency between the two subsets is examined. PWV is one of such environmental variables, and a null test between high-PWV and low-PWV subsets is often included in a test suite. The threshold of this split is PWV ∼ 1 mm for experiments in the Atacama Desert (Choi et al. 2020). Thus, PWV measurements with good fractional accuracy in the 1–3 mm range have a variety of uses in CMB analysis.

### 7.1 Evaluation of GNSS PWV for the data selection in CMB experiments

Here, we discuss the performance of GNSS PWV for CMB data analysis. The average time of PWV measurements is set to 1 h, which approximates the length of the basic unit of CMB data. As shown in Section 5, hourly GNSS PWV shows a statistical uncertainty of 0.52 mm and a systematic offset of −0.05 mm.

Table 2 shows how accurately we can judge whether PWV is less than 3, 2, and 1 mm using the GNSS data. For example, when data are split at PWV = 3 mm, GNSS PWV makes only a small percentage of error assuming that the radiometer PWV is the true value. Of the entire data, 1.7 per cent are misjudged as PWV$_{GNSS}$ > 3 mm while PWV$_{RM}$ < 3 mm, and 1.9 per cent are misjudged as PWV$_{GNSS}$ < 3 mm while PWV$_{RM}$ > 3 mm.

**Table 1.** An overview of other published GNSS PWV comparison studies at dry places where the mean annual PWV is less than 5 mm. The parameter $\Delta t$ is the average time of the GNSS PWV data; $a$ and $b$ are the linear fit parameters in equation (7); and STD is the standard deviation of the difference between the GNSS PWV and the compared PWV.

| Reference | Location | Compared method | Max PWV (mm) | $\Delta t$ (h) | $a$ | $b$ (mm) | STD (mm) |
|---|---|---|---|---|---|---|---|
| García-Lorenzo et al. (2010) | Canary Islands | Radiometer | <15 | 2 | 1.1 | 1.3 | 1.5 |
| Buehler et al. (2012) | Northern Sweden | Radiosonde | <25 | 2 | 0.965 | 0.357 | 0.66 |
| Negusini et al. (2021) | Antarctica | Radiosonde | 3–15 | 1 | 0.58–0.89 | −0.04–2.17 | 0.61–1.04 |
| This study | Atacama | Radiometer | 9.7 | 1 | 0.996 | −0.05 | 0.52 |

**Table 2.** The fractions of $PWV_{GNSS}$ with respect to $PWV_{RM}$ split by a threshold of 3, 2, and 1 mm.

|  | $PWV_{GNSS} < 3$ mm | $PWV_{GNSS} > 3$ mm |
|---|---|---|
| $PWV_{RM} < 3$ mm | 87.2 per cent | 1.7 per cent |
| $PWV_{RM} > 3$ mm | 1.9 per cent | 9.2 per cent |
|  | $PWV_{GNSS} < 2$ mm | $PWV_{GNSS} > 2$ mm |
| $PWV_{RM} < 2$ mm | 75.8 per cent | 3.2 per cent |
| $PWV_{RM} > 2$ mm | 3.0 per cent | 18.0 per cent |
|  | $PWV_{GNSS} < 1$ mm | $PWV_{GNSS} > 1$ mm |
| $PWV_{RM} < 1$ mm | 44.4 per cent | 9.8 per cent |
| $PWV_{RM} > 1$ mm | 7.3 per cent | 38.5 per cent |

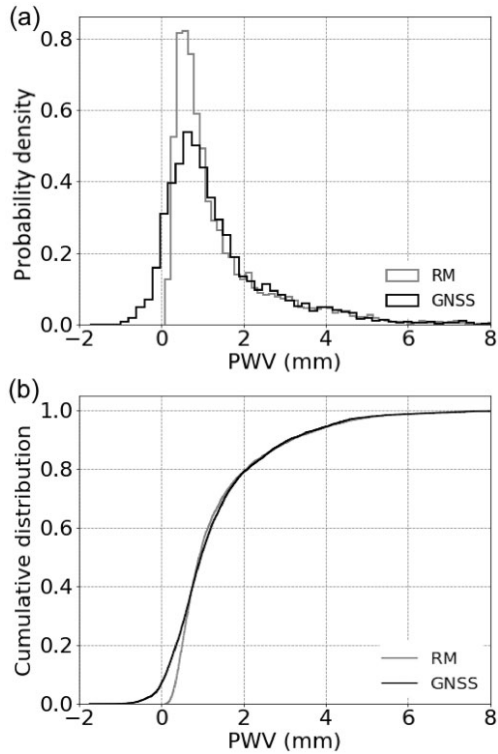

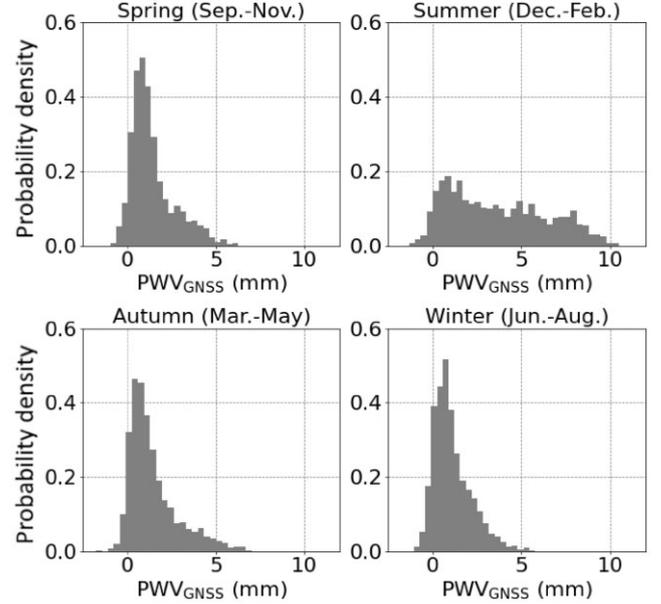

**Figure 8.** (a) Histograms of GNSS PWV and the radiometer PWV at the SA site from 2021 April to 2022 April. (b) Cumulative plots of GNSS PWV and radiometer PWV.

### 7.2 Annual PWV trend at Cerro Toco

Here, we demonstrate the site evaluation for the millimetre/submillimetre-wave observatories using GNSS data. The data range from 2021 April to 2022 April as in Section 5.

Fig. 8(a) shows the histograms of the GNSS PWV and radiometer PWV, and Fig. 8(b) shows their cumulative plots. GNSS estimates the PWV distribution with good accuracy where PWV is higher than 1 mm.

The seasonal distribution of GNSS PWV is shown in Fig. 9. The fractions of time when PWV is less than 3, 2, and 1 mm for each month are shown in Fig. 10. It is the advantage of GNSS instruments to continuously evaluate site conditions throughout the year.

**Figure 9.** Histograms of GNSS PWV at the SA site by season from 2021 April to 2022 April.

## 8 CONCLUSION

This study focused on how accurately PWV can be measured using GNSS data at Cerro Toco in the Atacama Desert to support millimetre/submillimetre-wave astronomical observations. We used CSRS-PPP to compute ZTD from GNSS data from 2021 April to 2022 April. ZHD is estimated using two methods, one using the co-located barometer data and the other using VMF1 outputs instead of the barometer. By comparing the GNSS PWV with the radiometer PWV, we evaluated the accuracy of the GNSS PWV.

GNSS PWV supported by the barometer data showed a small offset of −0.05 mm. Using the VMF1 outputs, GNSS PWV showed a larger offset. We assume that the offset strongly depends on the accuracy of ZHD estimation model. The statistical uncertainty of GNSS PWV is 0.52 mm with 1 h time average. The statistical uncertainty can be reduced by blocking multireflected GNSS signals, using multiple types of GNSS satellites (Li et al. 2015; Lu et al. 2017; Pan & Guo 2018; Bahadur 2022), and improving analysis methods (Calero-Rodríguez et al. 2023). GNSS instruments can continuously measure hourly PWV with better than 1 mm accuracy, even in occasional adverse weather conditions at Cerro Toco.

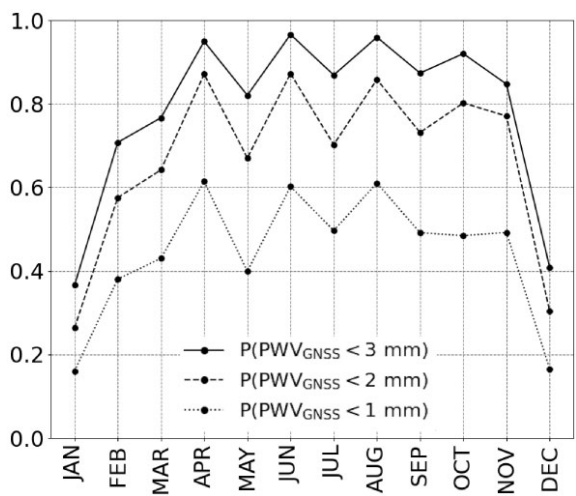

**Figure 10.** Fractions of GNSS PWV at the SA site less than 3, 2, and 1 mm for each month. The data are from 2021 April to 2022 April. SA site had good atmospheric conditions from April to November.

GNSS PWV data have a wide range of applications in support of CMB experiments, including site evaluation, data selection, instrument characterization, and null testing. GNSS instruments are reliable and easily accessible tools for PWV measurements and hold promise for improving future millimetre/submillimetre-wave observations.

**ACKNOWLEDGEMENTS**

The authors greatly appreciate the SA collaboration for supporting acquisition of the GNSS data and the NRCan for the CSRS-PPP online service. We also thank the Max-Planck-Institut fur Radioastronomie, the European Southern Observatory, and the Onsala Space Observatory for the APEX weather data. We also thank Nolberto Oyarce and José Cortes for the installation of the GNSS antenna and receiver system. This work was supported by JSPS KAKENHI Grant Number JP18H01240. JS would also like to thank JSR Fellowship, the University of Tokyo, and IGPEES, WINGS Programme, the University of Tokyo for supporting this research.

**DATA AVAILABILITY**

The GNSS data set supporting this study is available from the corresponding authors, J. Sugiyama and H. Nishino, upon reasonable requests. The APEX weather station and radiometer data are available at the APEX Weather Query Form (http://archive.eso.org/wdb/wdb/eso/meteo_apex/form).

# APPENDIX A: MODELLING THE REGIONAL WEIGHTED MEAN TEMPERATURE IN THE ATACAMA DESERT FROM RADIOMETER AND GNSS DATA

In this appendix, we discuss the derivation of the relationship between the weighted mean temperature $T_m$ and the surface temperature $T_s$ from our GNSS and radiometer data. $T_m$ is a meteorological parameter used in equation (5) to estimate GNSS PWV. It is calculated as

$$T_m = \frac{\int \frac{e(z)}{T(z)} dz}{\int \frac{e(z)}{T(z)^2} dz}, \quad (A1)$$

where $e(z)$ and $T(z)$ are the water vapour pressure and the temperature, respectively, as a function of zenith height $z$ (Davis et al. 1985). $T_m$ is usually estimated using radiosonde data of temperature and relative humidity (RH) at several altitudes. To enable the estimation of $T_m$ from the ground, the following empirical approximation is often used:

$$T_m = c \cdot T_s + d. \quad (A2)$$

**Table A1.** The regional $T_m$ models established by previous studies and this study.

| $T_m$ model | Region | Author |
|---|---|---|
| $T_m = 0.72 T_s + 70.2$ | US | Bevis et al. (1992) |
| $T_m = 1.07 T_s - 31.5$ | Taipei | Liou et al. (2001) |
| $T_m = 0.75 T_s + 63$ | India | Suresh Raju et al. (2007) |
| $T_m = 0.84 T_s + 48$ | Western Pacific | Suparta & Iskandar (2013) |
| $T_m = 0.62 T_s + 89.13$ | Antarctica | Negusini et al. (2016) |
| $T_m = 0.73 T_s + 69.7$ | Egypt | Elhaty et al. (2019) |
| $T_m = 1.15 T_s - 48.6$ | Atacama | This study |

The coefficients $c$ and $d$ are known to vary from region to region. A number of studies have established the regional $T_m$ models, some of which are shown in Table A1. In the arid regions, however, it is difficult to measure $T_m$ using radiosondes because the dry bias on the RH measurement is not negligible (see e.g. Otarola, Querel & Kerber 2011).

Here, we estimated the regional $T_m$ model at the SA site, located in the Atacama Desert, with the ground-based instruments of GNSS and the radiometer. First, GNSS ZWD was calculated as described in Section 4. Then, the data were divided into groups with respect to $T_s$ as shown in Fig. A1. The slope of the dotted line represents $1/\Pi$ for each $T_s$ bin. To estimate $\Pi$, we removed data points with RH greater than 40 per cent, which amount to 15 per cent of the entire data set. High humidity can cause bias in PWV measurements,[8] which is indeed seen in the top left panels in Fig. A1. Since $T_s$ and RH are highly correlated, bias in the small subset can lead to significant systematics here.

The radiometer PWV and GNSS ZWD were linearly fitted and $1/\Pi$ was obtained for each $T_s$ group. The relationship between $T_s$, $\Pi$, and $T_m$ is shown in Fig. A2. $\Pi$ is converted to $T_m$ using equation (5). The best-fitting model of our data set is

$$T_m = 1.15 T_s - 48.6. \quad (A3)$$

The fit parameters are $c = 1.15 \pm 0.31$ and $d = -48.6 \pm 84$, with correlation coefficient of $-1.0$. Our result is consistent with other regional $T_m$ models.

The fractional variation of $\Pi$ is given as

$$\frac{\sigma_\Pi}{\Pi} = \frac{k_3}{k_3 + k_2' T_m} \frac{\sigma_{T_m}}{T_m}. \quad (A4)$$

With typical values in Atacama, equation (A4) can be approximated as follows:

$$\frac{\sigma_\Pi}{\Pi} \sim 0.98 \frac{\sigma_{T_m}}{T_m}. \quad (A5)$$

The variation of $T_m$ in this study is about $\pm 10$ K, resulting in a $\Pi$ variation of $\sim 4$ per cent. Our PWV estimate in Atacama improves by a few per cent with the surface temperature data.

This ground-based method to estimate the $T_m$ model is easier to conduct than the radiosonde measurements; it is less expensive, and it is not affected by the dry bias. By continuing measurements for several years, the uncertainties of this $T_m$ model would decrease, potentially making the accuracy of this method competitive with that of other methods.

---

[8]For example, radiometer measurements are known to overestimate PWV in the presence of snow or liquid water (Shangguan et al. 2015).

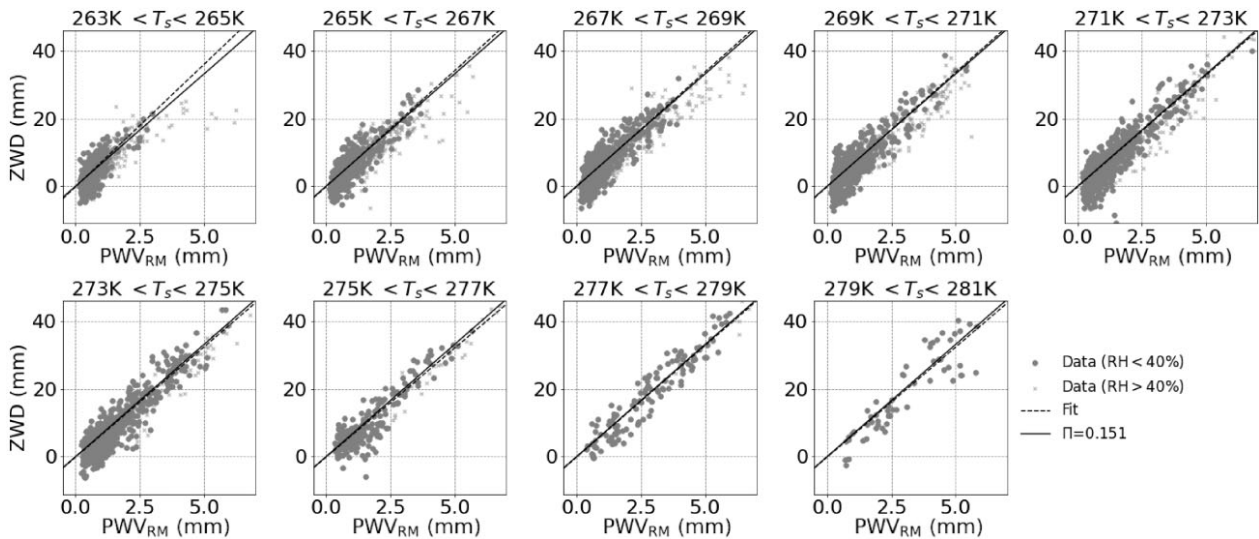

**Figure A1.** Scatter plot of hourly PWV versus ZWD for each surface temperature range in the Atacama Desert. The dashed line corresponds to the linear fit and the solid line corresponds to $\Pi = 0.151$. PWV was measured by the radiometer (Section 4.3) and ZWD was measured by the GNSS instruments. In the linear fit, data with RH greater than 40 per cent are removed to mitigate the radiometer systematics due to the possible presence of liquid water or snow.

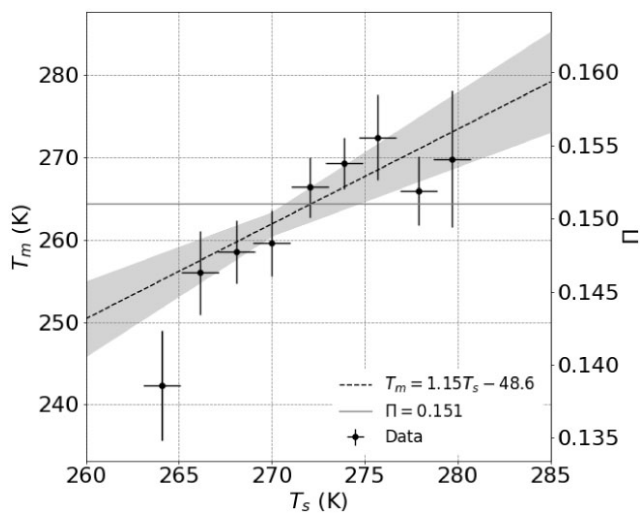

**Figure A2.** The relationship between the weighted mean temperature $T_m$ and the surface temperature $T_s$ in the Atacama Desert obtained from Fig. A1.

This paper has been typeset from a TeX/LaTeX file prepared by the author.